\documentclass{moriond}

\def\Journal#1#2#3#4{{#1} {\bf #2}, #3 (#4)}

\def\EPJC{{\em Eur. Phys. J.} C}
\def\JHEP{\em JHEP}
\def\JPG{{\em J. Phys.} G}

\def\NIMA{{\em Nucl. Instrum. Methods} A}
\def\NPB{{\em Nucl. Phys.} B}

\def\PRL{\em Phys. Rev. Lett.}
\def\PRD{{\em Phys. Rev.} D}
\def\PTP{\em Prog. Theor. Phys.}

\def\be{\begin{equation}}
\def\ee{\end{equation}}
\def\bea{\begin{eqnarray}}
\def\eea{\end{eqnarray}}

\input{definitions}

\newcommand{\Photo}{}

\begin{document}
\vspace*{4cm}
\title{Graph Neural Network Flavor Tagger and measurement of $\sin2\beta$ at Belle~II}

\author{ P.~Stavroulakis on behalf of the Belle II collaboration }

\address{Universit\'e de Strasbourg, CNRS, IPHC UMR 7178, F-67000 Strasbourg, France}

\maketitle\abstracts{
We present GFlaT, a new algorithm that uses a graph-neural-network to determine the flavor of neutral \PB mesons produced in \FourS decays. We evaluate its performance using \PB decays to flavor-specific hadronic final states reconstructed in a \SI{362}{fb^{-1}} sample of electron-positron collisions recorded at the \FourS resonance with the Belle II detector at the SuperKEKB collider. We achieve an effective tagging efficiency of \SI{37.40 +- 0.43 +- 0.36}{\percent}, where the first uncertainty is statistical and the second systematic, which is 18\% better than the previous Belle II algorithm. Demonstrating the algorithm, we use $\PBzero\to\PJpsi\PKshortzero$ decays to measure the direct and mixing-induced \CP violation parameters, \CCP = \SI{-0.035 +- 0.026 +- 0.013}{} and \SCP = \SI{0.724 +- 0.035 +- 0.014}{}, from which we obtain $\beta$ = \SI{23.2 +- 1.5 +- 0.6}{\degree}.
}

\section{Introduction}\label{sec:intro}

In the standard model of particle physics, an irreducible complex phase in the Cabibbo-Kobayashi-Maskawa (CKM) matrix~\cite{Kobayashi:1973fv} gives rise to \CP violation.
Measurements of mixing-induced \CP violation in \PBzero meson decays constrain the values of the CKM-unitarity-triangle angles $\beta$ and $\alpha$,\footnote{These angles are also known as $\phi_1$ and $\phi_2$.} which helps probe for sources of \CP violation beyond the standard model.
Such measurements require knowledge of the neutral \PB meson flavor. At \belletwo~\cite{Abe:2010gxa}, \PBzero and \APBzero mesons are produced in coherent pairs from $\APelectron\Pelectron$ collisions at the \FourS resonance.
Since they are entangled, tagging the flavor of one of the mesons, \PBtag, at the time of its decay determines the flavor of the other one, \PBsig, at that same time~\cite{Bigi:1981qs}. The probability density to observe \PBsig decay to a \CP eigenstate at a time $\Delta t$ from when \PBtag decays with flavor \qtag ($+1$ for \PBzero, $-1$ for \APBzero) is given by

\begin{equation}
    P(\Delta t, \qtag) = \frac{e^{-\abs*{\Delta t}/\tau}}{4\tau} \big\{1 + \qtag (1 - 2w)[\SCP \sine(\Delta m_{\Pdown} \Delta t) - \CCP \cosine(\Delta m_{\Pdown} \Delta t)]\big\},
    \label{eq:dt_cp_perf}
\end{equation}
where \qtag is determined by the flavor tagger, $w$ is the probability to wrongly determine it, $\tau$ is the \PBzero lifetime, and $\Delta m_{\Pdown}$ is the difference of masses of the \PBzero mass eigenstates~\cite{hadpaper}. Here \SCP and \CCP are parameters that quantify mixing-induced and direct \CP violation. In the standard model, $\SCP = \sin2\beta$ and $\CCP = 0$ to good precision~\cite{DeBruyn:2014oga,Barel:2020jvf}.

In order to be able to measure \SCP and \CCP as accurately as possible, one needs to have precise knowledge of $w$. We determine it from events where \PBsig decays to $\PBzero\to\PDoptstarminus\Ppiplus$, for which
\begin{equation}
    P(\Delta t, \qsig, \qtag) = \frac{e^{-\abs*{\Delta t} / \tau}}{4\tau} \qty{1 - \qsig\qtag (1 - 2w) \cosine(\Delta m_{d} \Delta t)},
    \label{eq:dt_flav_perf}
\end{equation}

Besides flavor tagging effects, the finite resolution of the detector also affects the determination of $\Delta t$ and therefore all time-dependent measurements. We account for this by convolving Eq.~\ref{eq:dt_cp_perf} and \ref{eq:dt_flav_perf} with the $\Delta t$ resolution model used in Ref~\cite{hadpaper}.

\section{Flavor Tagging at \belletwo}\label{sec:flavor tagging}

\subsection{Category-based Flavor Tagger}\label{ssec:cat-based tagger}
Until recently, flavor tagging was performed at \belletwo via a category-based approach~\cite{Belle-II:2021zvj} which uses topological, kinematic and particle identification information from the decay of the \PBtag in order to assign it to one of several flavor-specific tagging categories. The assignment with the highest probability is then used to make a prediction on the flavor of \PBtag, exploiting the correlation between the charge of one of the final state particles and the flavor of the decaying \PB meson. The output of this category-based flavor tagger consists of the flavor of \PBtag at the time of its decay, \qtag, and the confidence of the flavor assignment, $r$, which can range anywhere between 0 (lowest confidence) to 1 (highest confidence).

The mistag probability, $w$, is characteristic of the flavor tagger and is usually determined in seven bins of $r$ defined by the edges $[0.0, 0.1, 0.25, 0.45, 0.6, 0.725, 0.875, 1.0]$. To evaluate the performance of the flavor tagger we use the effective tagging efficiency,
\begin{equation}
    \efftag = \sum_i \effi (1 - 2 w_i)^2,
\end{equation}
where \effi is the efficiency to assign a flavor to \PBtag and $w_i$ is the mistag probability in the $i$-th $r$ bin. A higher effective tagging efficiency directly translates to more statistically precise measurements of \SCP and \CCP.

\subsection{Graph-Neural-Network Flavor Tagger}\label{ssec:gflat}
Recently, a new flavor tagging algorithm~\cite{gflat} based on a dynamic-graph-convolutional neural network~\cite{Qu:2019gqs} was developed at \belletwo, which makes use of the relational information between the final state particles in the tag-side in order to increase the effective tagging efficiency.

\begin{figure}[tb]
    \centering
    \includegraphics[width=0.5\linewidth]{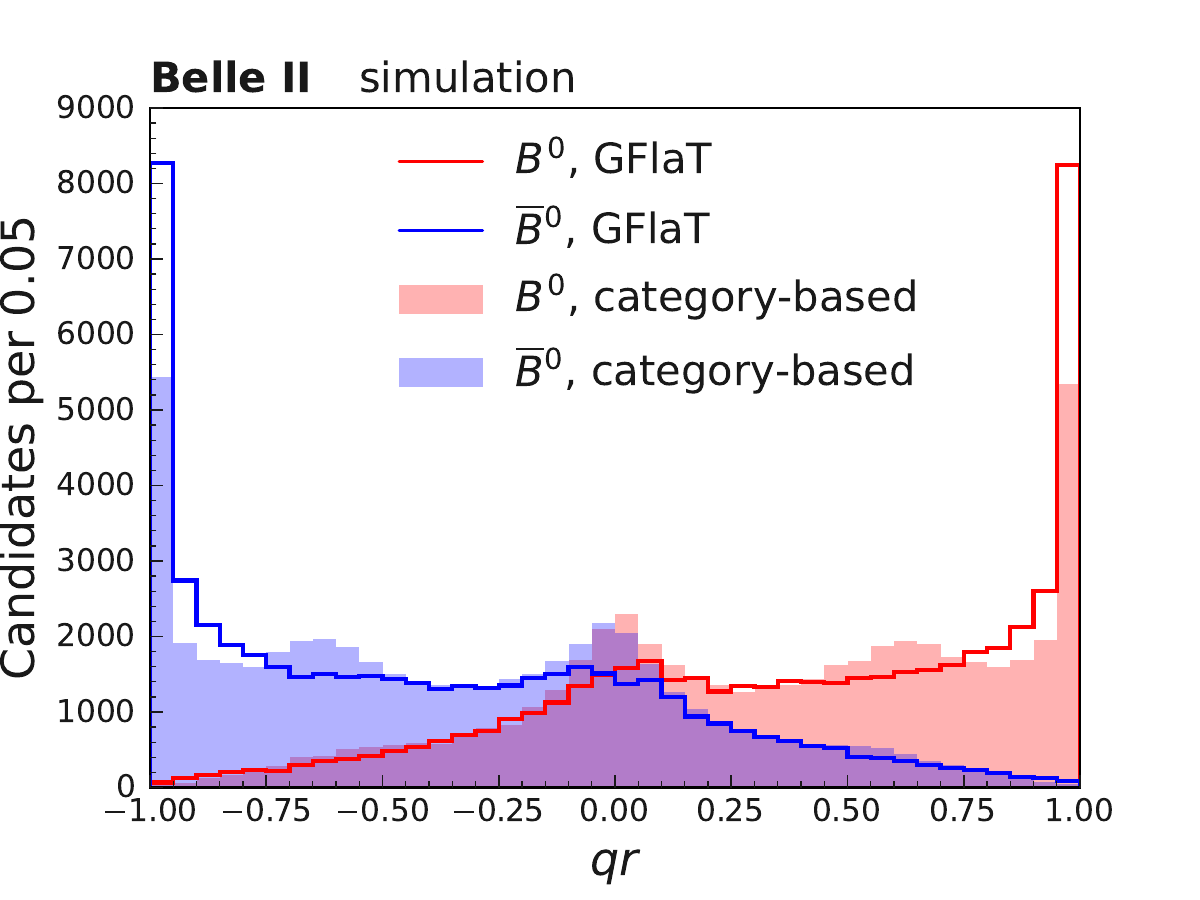}
    \caption{Distributions of $qr$ for true \PBzero and \APBzero from \gflat and the category-based flavor tagger in simulated data.}
    \label{fig:qr_comparison}
\end{figure}

Figure~\ref{fig:qr_comparison} shows the $qr$ distributions of both the category-based and graph-neural-network-based flavor taggers. The latter improves on the former via better tagging of events in which the \PBtag decays into final states without any charged leptons. This leads to a $qr$ distribution that is more peaking at the edges and flatter in the middle and, consequently, to a relative improvement of $20\%$ in effective tagging efficiency, that is estimated from simulation.

\section{Calibration of \gflat}\label{sec:calib}

We use $\PBzero\to\PDoptstarminus\Ppiplus$ decays to evaluate the performance of \gflat and calibrate its parameters, since they allow to access the flavor of \PBsig without the need for flavor tagging. To achieve this, we perform a time-dependent analysis, that allows to also extract the parameters of the $\Delta t$ resolution function.

We specifically reconstruct $\PDminus\to\PKplus\Ppiminus\Ppiminus$ decays and $\PDstarminus\to\APDzero\Ppiminus$ decays with $\APDzero\to\PKplus\Ppiminus$, $\PKplus\Ppiminus\Ppizero$, or $\PKplus\Ppiminus\Ppiplus\Ppiminus$, using charged particle tracks and calorimeter clusters. \PBzero candidates are reconstructed from a \PDoptstarminus and a charged particle track consistent with a \Ppiplus. A kinematic fit is performed for each \PBzero candidate in order to determine its decay vertex position.

We first perform an unbinned maximum likelihood fit to the $\Delta E$ distribution, defined as $\Delta E \equiv E - E\Sub{beam}$, where $E\Sub{beam}$, $E$ are the beam energy and \PBzero energy in the c.m.~frame, respectively. We then subtract the background in the $\Delta t$ distribution using \sWeight~\cite{Pivk:2004ty} and fit the resulting signal distribution simultaneously in the 7 bins of $r$, 2 flavors of \PBtag and 2 flavors of \PBsig to extract the parameters of the flavor tagger and $\Delta t$ resolution function.
The parameters obtained with the procedure described above are considered~\footnote{This assumption was verified between $\PBzero\to\PDoptstarminus\Ppiplus$ and $\PBzero\to\PJpsi\PKshortzero$ decays in simulation.} to be independent of decay mode and are used in all subsequent time-dependent measurements.

\begin{figure}[htbp]
    \centering
    \begin{minipage}{0.55\textwidth}
        \centering
        \includegraphics[width=\textwidth]{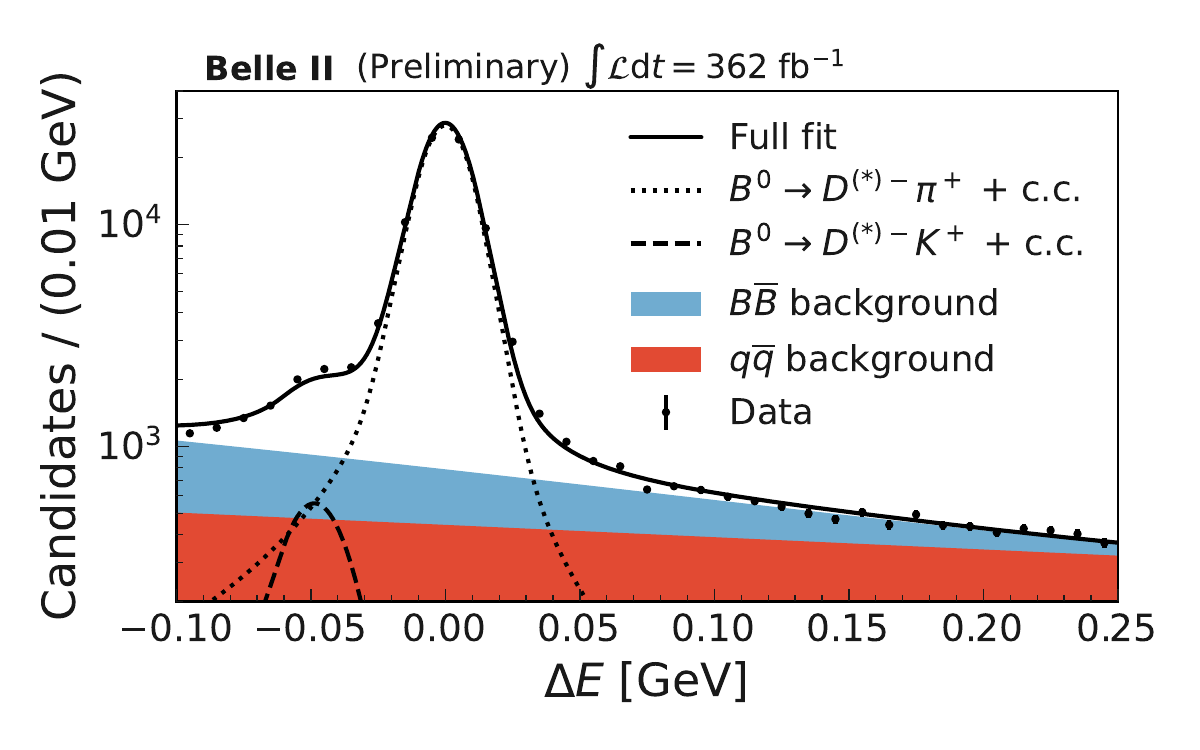}
    \end{minipage}\hfill
    \begin{minipage}{0.45\textwidth}
        \centering
        \includegraphics[width=\textwidth]{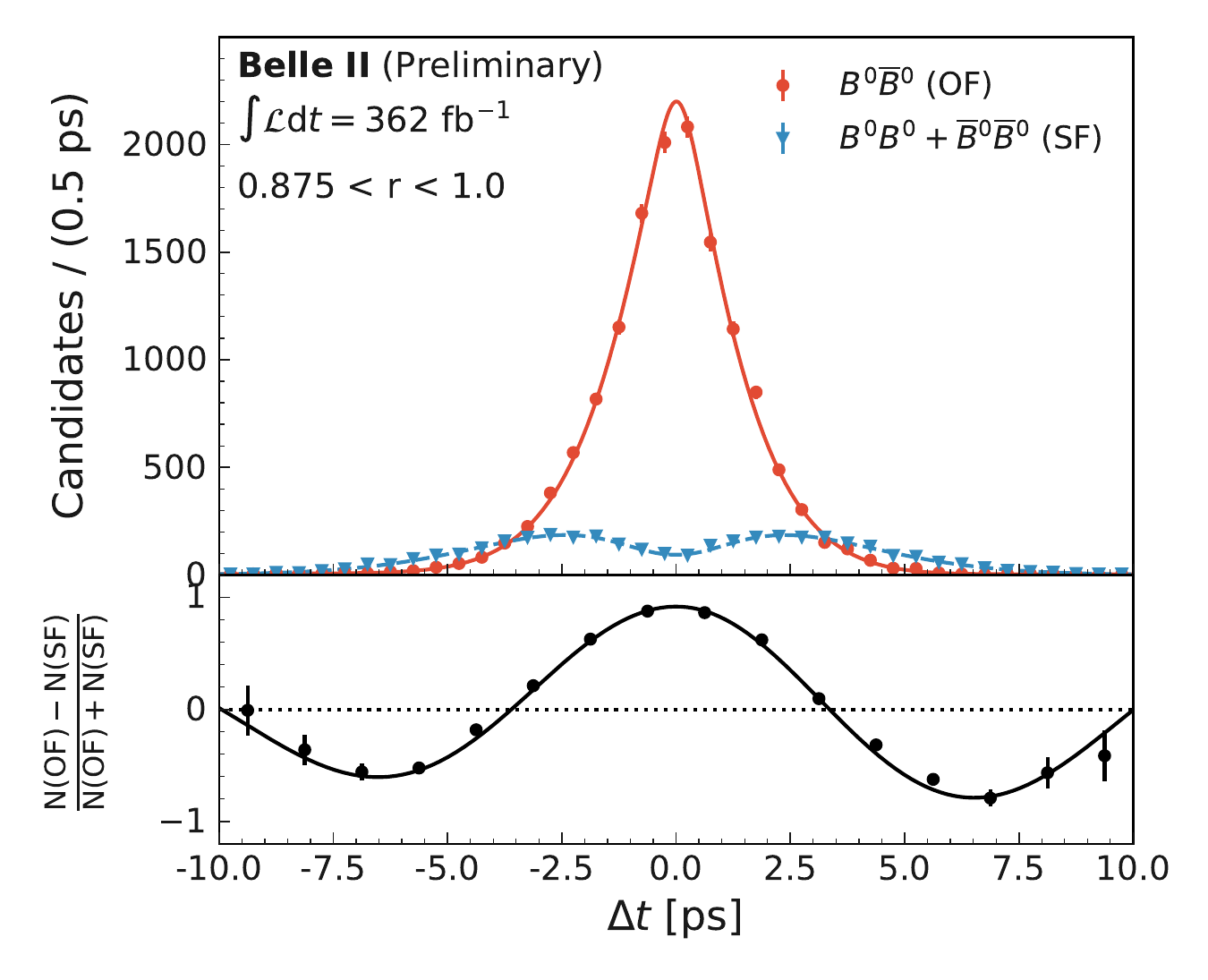}
    \end{minipage}
    \caption{Distributions of $\Delta E$~(left) and background-subtracted $\Delta t$~(right) distributions in the seventh $r$-bin for $\PBzero\to\PDoptstarminus\Ppiplus$~(points) and the best-fit functions~(lines). The background components of the fit to the $\Delta E$ distribution are stacked.}
    \label{fig:hadcalib_fit}
\end{figure}

Figure~\ref{fig:hadcalib_fit} shows the signal extraction fit to the $\Delta E$ distribution as well as the fit to the background-free $\Delta t$ distribution in the $r$-bin with the highest confidence.
We obtain an effective tagging efficiency of
\begin{equation*}
    \efftag = (37.40 \pm 0.43 \pm 0.36)\%
\end{equation*}

For comparison, we evaluate the effective tagging efficiency using the category-based flavor tagger on the same set of data and obtain $\efftag = (31.68 \pm 0.45)\%$. Therefore, we conclude that \gflat yields a relative improvement of 18\% in effective tagging efficiency over the category-based flavor tagger.

\section{Measurement of $\sin2\beta$ using \gflat}\label{sec:sin2b}

We validated \gflat through a measurement of \SCP and \CCP in $\PBzero\to\PJpsi\PKshortzero$ decays.
We reconstruct \PJpsi candidates via $\PJpsi\to\APelectron\Pelectron$ or $\APmuon\Pmuon$ and \PKshortzero candidates via $\PKshortzero\to\Ppiplus\Ppiminus$.

\begin{figure}[htbp]
    \centering
    \begin{minipage}{0.55\textwidth}
        \centering
        \includegraphics[width=\textwidth]{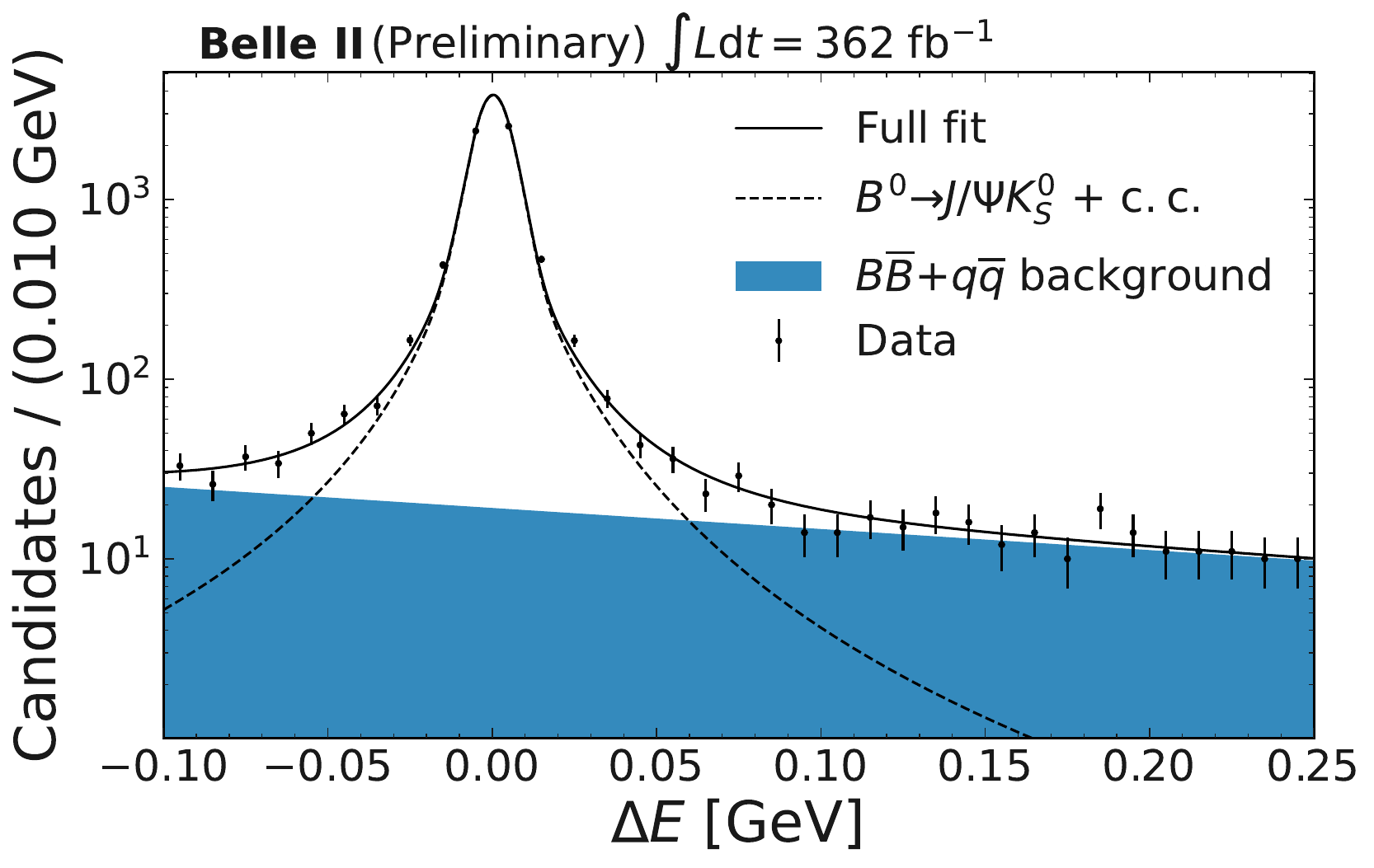}
    \end{minipage}\hfill
    \begin{minipage}{0.4\textwidth}
        \centering
        \includegraphics[width=\textwidth]{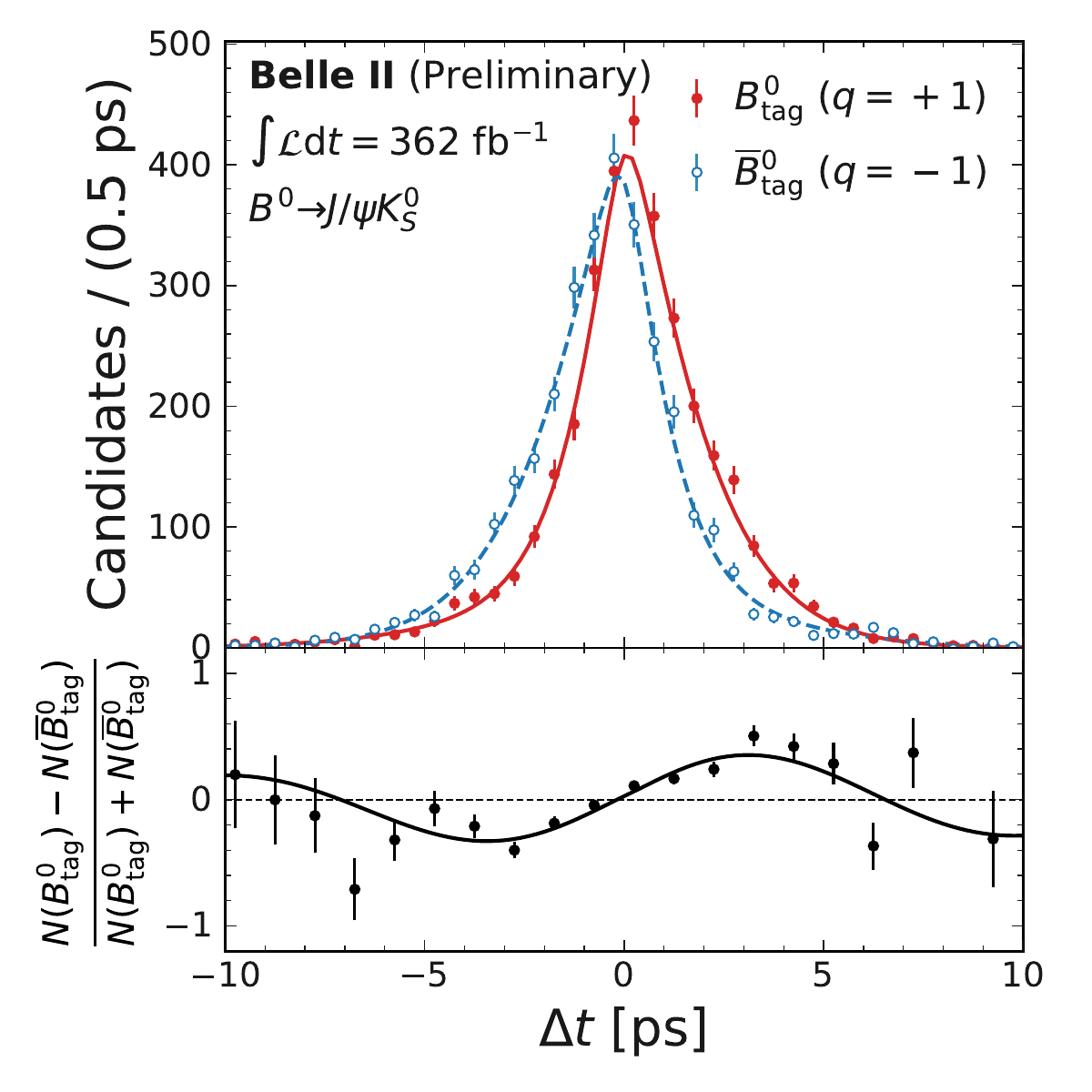}
    \end{minipage}
    \caption{Distributions of $\Delta E$~(left) and background-subtracted $\Delta t$~(right) distributions in the full $r$ range for $\PBzero\to\PJpsi\PKshortzero$~(points) and the best-fit functions~(lines).}
    \label{fig:KS_defit_realdata}
\end{figure}

A signal extraction fit is performed on the $\Delta E$ distribution, from which $sWeights$ are computed and then used to subtract the background in the $\Delta t$ distribution. We then fit the resulting $\Delta t$ distribution simultaneously in the 7 bins of $r$ and 2 \PBtag flavors to extract \SCP and \CCP. For this fit, we use the $\Delta t$ resolution model and flavor tagging parameters that were calibrated using $\PBzero\to\PDoptstarminus\Ppiplus$ decays. The fit procedure is validated using $\PBzero\to\PJpsi\PKstarzero$ decays, where both \SCP and \CCP are expected to be zero.
Figure~\ref{fig:KS_defit_realdata} shows the fits to the $\Delta E$ and background-subtracted $\Delta t$ distributions in events containing $\PBzero\to\PJpsi\PKshortzero$ decays.
We obtain
\begin{alignat*}{5}
      \SCP &=~ && &  0.724 && ~\pm~ 0.035 && ~\pm~ 0.014, \\ 
      \CCP &=~ && & -0.035 && ~\pm~ 0.026 && ~\pm~ 0.013,
\end{alignat*}
The statistical uncertainties are 8\% and 7\% smaller, respectively, than they would be if measured using the category-based flavor tagger, which is expected given the higher effective tagging efficiency of \gflat.
From \SCP, we calculate $\beta$ = \SI{23.2 +- 1.5 +- 0.6}{\degree}.\footnote{The other solution $\pi/2-\beta$ is excluded from independent measurements~\cite{BaBar:2015oxm}} which agrees with previous measurements by \babar~\cite{BaBar:2009byl}, \belle~\cite{Belle:2012paq} and \lhcb~\cite{LHCb:2023zcp}.

\section*{References}

\begin{thebibliography}{99}

\bibitem{Kobayashi:1973fv}M. Kobayashi and T. Maskawa, \Journal{\PTP}{49}{652}{1973}.

\bibitem{Abe:2010gxa}T. Abe et al. (Belle II Collaboration), arXiV:physics/1011.0352.

\bibitem{Bigi:1981qs}I. I. Y. Bigi and A. I. Sanda, \Journal{\NPB}{193}{85-108}{1981}.

\bibitem{hadpaper}F. Abudin\'en et al. (Belle II Collaboration), \Journal{\PRD}{107}{091102}{2023}.

\bibitem{DeBruyn:2014oga}K. De Bruyn and R. Fleischer, \Journal{\JHEP}{03}{145}{2015}.

\bibitem{Barel:2020jvf}M. Barel and K. De Bruyn and R. Fleischer and E. Malami, \Journal{\JPG}{48}{065002}{2021}.

\bibitem{Belle-II:2021zvj}F. Abudin\'en et al. (Belle II Collaboration), \Journal{\EPJC}{82}{283}{2022}.

\bibitem{gflat}I. Adachi et al. (Belle II Collaboration), arXiV:physics/2402.17260.

\bibitem{Qu:2019gqs}H. Qu and L. Gouskos, \Journal{\PRD}{101}{056019}{2020}.

\bibitem{Pivk:2004ty}M. Pivk and F. Le Diberder, \Journal{\NIMA}{555}{356-369}{2005}.

\bibitem{BaBar:2015oxm}A. Abdesselam et al. (BABAR Collaboration), \Journal{\PRL}{115}{121604}{2015}.

\bibitem{BaBar:2009byl}B. Aubert et al. (BABAR Collaboration), \Journal{\PRD}{79}{072009}{2009}

\bibitem{Belle:2012paq}I. Adachi et al. (Belle II Collaboration), \Journal{\PRL}{108}{171802}{2012}

\bibitem{LHCb:2023zcp}R. Aaij et al. (LHCb Collaboration), \Journal{\PRL}{132}{021801}{2024}

\end{thebibliography}
\end{document}